\documentclass[aps,twocolumn,floatfix,groupedaddress,amsmath,amssymb,prb]{revtex4}
\usepackage{graphicx}
\usepackage{mathrsfs}
\usepackage{slashbox}
\usepackage{array}
\usepackage{float}


\newcommand{\ket} [1] {| #1 \rangle}
\newcommand{\bra} [1] {\langle #1 |}

\makeatletter

\newcommand{\Rmnum}[1]{\expandafter\@slowromancap\romannumeral #1@}

\newcommand{\eref}[1]{Eq.~(\ref{#1})}

\newcommand{\fref}[1]{Fig.~\ref{#1}}
\newcommand{\lattice}{\mathcal{L}}

\begin{document}
\title{Entanglement renormalization and symmetry fractionalization}
\author{Sukhbinder Singh$^{1}$} 
\author{Nathan McMahon$^{2,3}$}
\author{Gavin Brennen$^{2}$}
\affiliation{$^{1}$ Max-Planck Institute for Gravitational Physics (Albert Einstein Institute), Potsdam, Germany}
\affiliation{$^{2}$Center for Engineered Quantum Systems, Dept. of Physics \& Astronomy, Macquarie University, 2109 NSW, Australia}
\affiliation{$^{3}$Center for Engineered Quantum Systems, School of Mathematics and Physics,
The University of Queensland, St Lucia, Queensland 4072, Australia}

\begin{abstract}
It is well known that the matrix product state (MPS) description of a gapped ground state with a global on-site symmetry can exhibit `symmetry fractionalization'. Namely, even though the symmetry acts as a \textit{linear} representation on the physical degrees of freedom, the MPS matrices---which act on some virtual degrees of freedom---can transform under a \textit{projective} representation. This was instrumental in classifying gapped symmetry protected phases that manifest in one dimensional quantum many-body systems. 
Here we consider the multi-scale entanglement renormalization ansatz (MERA) description of 1D ground states that have global on-site symmetries. We show that, in contrast to the MPS, the symmetry \textit{does not} fractionalize in the MERA description if the ground state is \textit{gapped}, assuming that the MERA preserves the symmetry at all length scales. However, it is still possible that the symmetry can fractionalize in the MERA if the ground state is \textit{critical}, which may be relevant for characterizing critical symmetry protected phases. Our results also motivate the presumed use of symmetric tensors to implement global on-site symmetries in MERA algorithms.
\end{abstract}

\maketitle

\section{Introduction} 
\begin{figure}[t]
  \includegraphics[width=7.5cm]{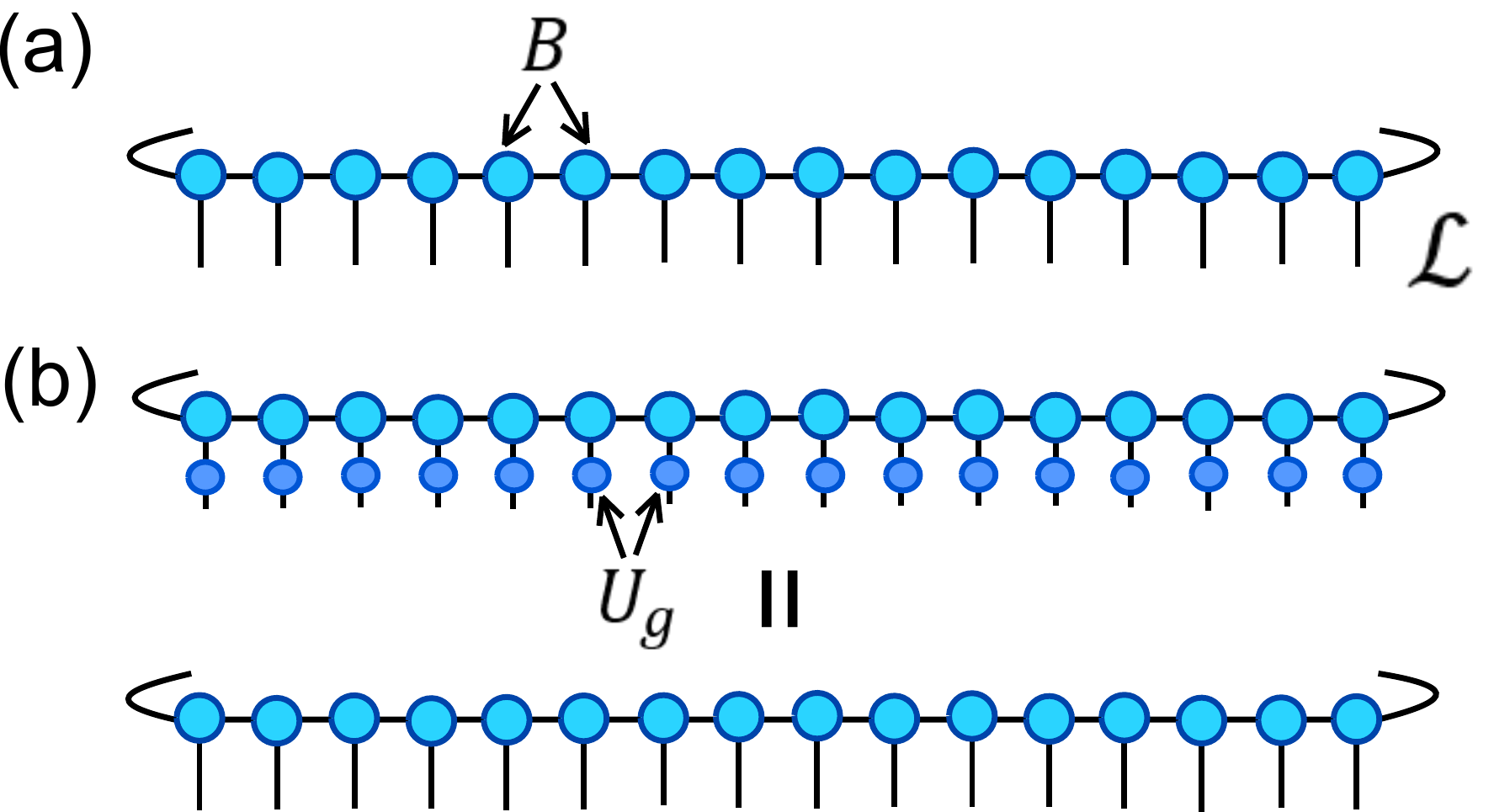}
\caption{\label{fig:mps} (a) A tensor network made of 3-index tensors arranged on a circle, which can be contracted to obtain the probability amplitudes of a matrix product state (see also Appendix \ref{sec:appendixA}). Open indices correspond to sites of a lattice $\mathcal{L}$. We will consider only translation-invariant MPS, which consists of copies of the same tensor $B$ everywhere. (b) An MPS with a global on-site symmetry $\mathcal{G}$. Here $U_g$ is a unitary \textit{linear} representation of the symmetry on each lattice site. Namely, the representation fufills the group product exactly, $U_gU_h=U_{gh}$ for all group elements $g,h$ of $\mathcal{G}$.}
\end{figure}

Characterizing symmetries in tensor network states has recently played an instrumental role in the classification of gapped quantum phases of matter. For example, on a one dimensional lattice, ground states of gapped local Hamiltonians can be efficiently described as \textit{matrix product states} (MPS), whose probability amplitudes are obtained by contracting a tensor network such as the one illustrated in \fref{fig:mps}(a) \cite{mps1,mps2}. Consider an MPS description of a 1D gapped ground state with a global on-site symmetry $\mathcal{G}$, such that the symmetry acts as a linear representation on each site, see \fref{fig:mps}(b).
If the MPS is also \textit{normal}\cite{mps3} (see also Appendix \ref{sec:appendixA}) then the MPS tensor $B$ fulfills, up to certain gauge transformations\cite{gauge},
\begin{equation}\label{eq:mpsSym}
\vcenter{\hbox{\includegraphics[width=3.5cm]{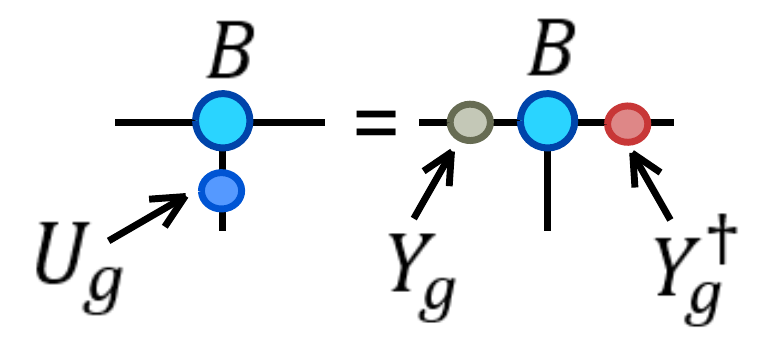}}}.
\end{equation}
Here, even though the symmetry acts \textit{linearly} on the physical sites, the matrices $Y_g$ can form a \textit{projective} representation of the symmetry\cite{MPSSym,MPSSym2}, which fulfills the group product only up to a phase as $Y_gY_h=e^{i\omega(g,h)}Y_{gh}$.\cite{cocycle}
This is sometimes referred to as `symmetry fractionalization' in the MPS. (If the MPS is also in a certain \textit{canonical form}\cite{mps3}, the matrices $Y_g$ are unitary.) If two (normal) symmetric MPSs carry inequivalent\cite{projrep} projective `bond representations' $Y_g$ and $Y'_g$ respectively, then they belong to different quantum phases protected by the same symmetry $\mathcal{G}$.\cite{CohmologyClassification,CohmologyClassification1,CohmologyClassification2, phaseDetection, phaseDetection1}  Thus, the MPS description of ground states characterizes different quantum phases in one dimension.

Notice that \eref{eq:mpsSym} is also a \textit{local} constraint---i.e., fulfilled by individual tensors---that results from imposing a \textit{global} symmetry on the total MPS tensor network. Besides setting the stage for possible symmetry fractionalization, this local realization of the global symmetry is also convenient in MPS simulations, where one may be interested in protecting the symmetry against numerical errors. Tensor $B$, which fulfills \eref{eq:mpsSym}, is an example of a \textit{symmetric tensor}, which loosely speaking is a tensor that commutes with the symmetry\cite{symTensorsProj}. It turns out that symmetric tensors have a sparse structure, which is determined by the representation theory of the symmetry\cite{symmetricTensors}. The sparse structure is often exploited to reduce computational costs in MPS simulations, while also protecting the symmetry.\cite{MPSSymImplement}

In this paper, we characterize symmetries in another prominent and efficient tensor network description of 1D ground states---the multi-scale entanglement renormalization ansatz\cite{ER, MERA} (MERA). Unlike the MPS, the MERA is also suitable for describing \textit{critical} ground states\cite{criticalMERA}. Inspired by the MPS, global on-site symmetries have also been implemented in the MERA by making judicious use of symmetric tensors. However, there is no formal proof that a global on-site symmetry \textit{necessarily} implies that the MERA tensors must be symmetric. That is, there is no MERA analog of \eref{eq:mpsSym}.

Despite this, the use of symmetric tensors has played a central role in several MERA applications, e.g., (i) targetting specific symmetry sectors and reducing computational costs in MERA simulations while exactly protecting the symmetry against numerical errors\cite{symmera}, (ii) determining non-local scaling and topological defect operators of conformal field theories that have a global symmetry\cite{nonlocalMERA},  (iii) building exact MERA descriptions of RG fixed points in 1D symmetry protected phases\cite{symProtectedMERA}, and (iv) realizing the bulk gauging of a global boundary symmetry\cite{symProtectedMERA,holospin}  in a holographic interpretation of the MERA\cite{Swingle}. 

In this paper, we first show that (under reasonable assumptions) global on-site symmetries in the MERA also necessarily lead to symmetric tensors (again, up to gauge transformations). More precisely, we show that if a MERA has a global on-site symmetry and the symmetry is protected at all length scales (as described in the next section), then one can always make its tensors symmetric by introducing suitable gauge transformations\cite{optimality}.

This, for one, motivates the presumed use of symmetric tensors for implementing global on-site symmetries in MERA algorithms e.g. in the context of applications (i)-(iv) listed above.
Second, it prompts the question: Can the symmetry fractionalize in the MERA description of a 1D ground state with a global on-site symmetry? (Since symmetric tensors are a prerequisite for symmetry fractionalization in the MPS.)
We will argue that, in contrast to the MPS, the answer is negative for a \textit{gapped} ground state. However, our argument does not apply to critical ground states. Thus, in the absence of any other restrictions, it is possible that the symmetry can still fractionalize in the MERA description of symmetric critical ground states. We suggest the possibility that inequivalent\cite{projrep} symmetry-fractionalized MERA states belong to different \textit{critical} symmetry protected phases, for example, see Refs.~\onlinecite{symProtectedCritical,symProtectedCritical1,AnomaliesMERA}.

\begin{figure}[t]
  \includegraphics[width=8cm]{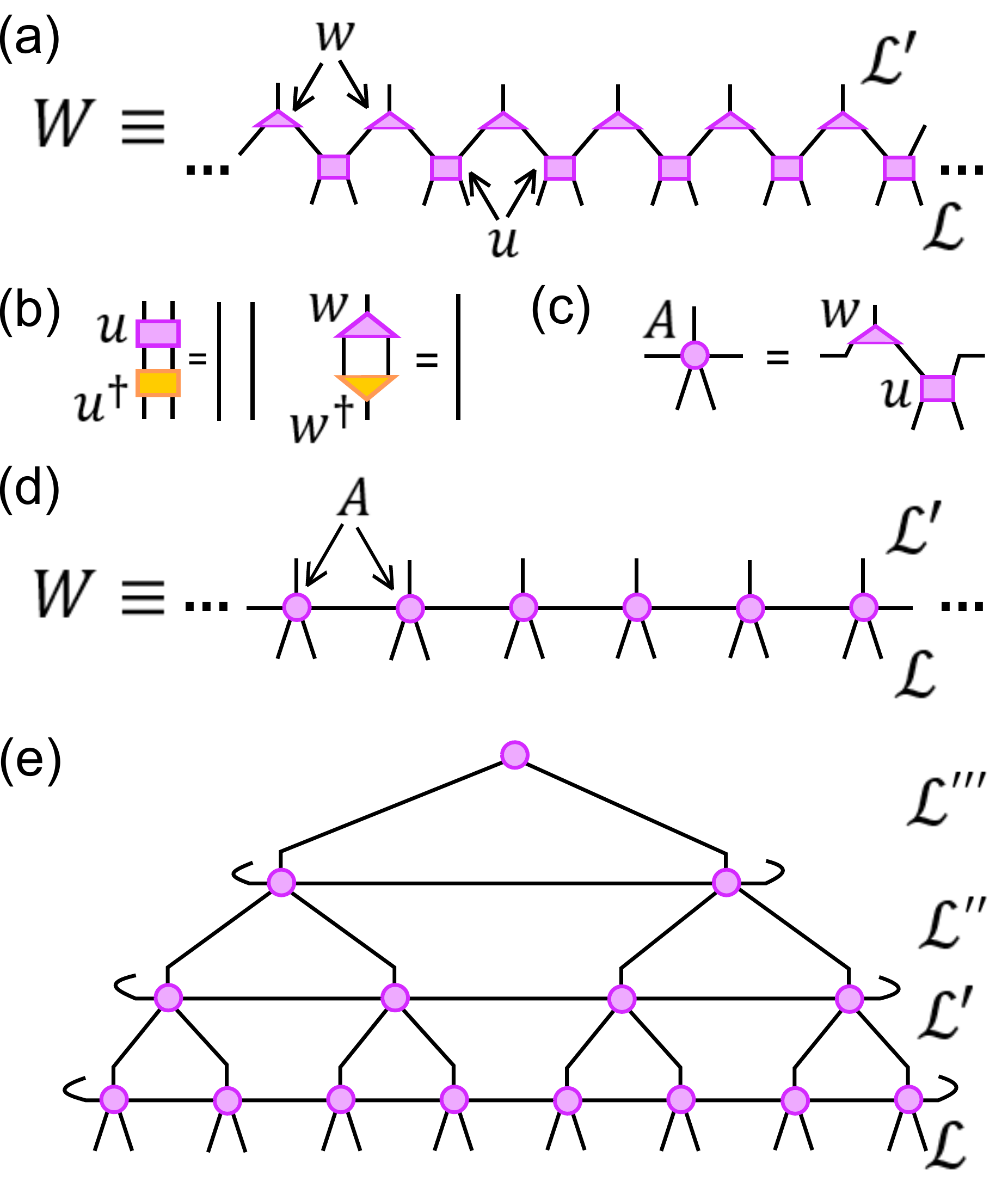}
\caption{\label{fig:ER} (a) An entanglement renormalization transformation $W$ as a tensor network that implements a linear map from a lattice $\lattice$ to a coarse-grained lattice $\lattice'$. For simplicity, we assume that copies of the same two tensors $u,w$ appear everywhere. (b) Tensors $u$ and $w$ are isometries, thus they fulfill these equalities. (c) Tensor $A$ is obtained by contracting $u$ and $w$.  (d) Entanglement renormalization transformation made instead from copies of tensor $A$. (e) MERA tensor network for a quantum many-body state of a lattice of 16 sites, obtained by composing several $W$ transformations, different tensors may appear in each $W$.}
\end{figure}

\section{MERA and symmetric tensors}
From the outset it is apparent that the MERA tensor network is quite different from the MPS. (In particular, this means that the proof of \eref{eq:mpsSym}, which relied on the specific structure of the MPS tensor network, cannot be applied directly to the case of the MERA.) For example, while the MPS is a 1D tensor network, the MERA extends in two dimensions. 

The extra dimension in the MERA can be understood as a length scale, since the MERA is generated by a real space renormalization group (RG) transformation, known as entanglement renormalization.\cite{ER} This RG transformation acts on the lattice by removing entanglement between blocks of sites before coarse-graining them. It can be described by a tensor network composed from isometric tensors $u$ and $w$ which represent the disentangling and coarse-graining components respectively, see \fref{fig:ER}(a,b). For convenience, we will contract tensors $u$ and $w$ to obtain tensor $A$, as shown in \fref{fig:ER}(c). The resulting tensor network, comprised of copies of $A$, is a (translation-invariant) \textit{matrix product operator}---the operator analog of an MPS---which implements a linear map $W$ from a fine-grained lattice $\mathcal{L}$ to coarse-grained lattice $\mathcal{L}'$, see \fref{fig:ER}(d).

Entanglement renormalization is evidently capable of generating RG flows with proper fixed points (in the thermodynamic limit) in 1D---both in gapped\cite{symProtectedMERA,symProtectedMERA1} and critical\cite{criticalMERA} systems, and also in 2D quantum systems with topological order\cite{topoMERA}. 

The MERA is a striped tensor network that is generated by composing several entanglement renormalization transformations, as illustrated in \fref{fig:ER}(e). The MERA description of a ground state can be interpreted as the RG flow of the ground state: discarding bottom stripes of the MERA yields a description of the ground state on a sequence of increasingly coarse-grained lattices $\mathcal{L} \rightarrow \mathcal{L}' \rightarrow \mathcal{L}'' \cdots$. 

\begin{figure}[t]
  \includegraphics[width=\columnwidth]{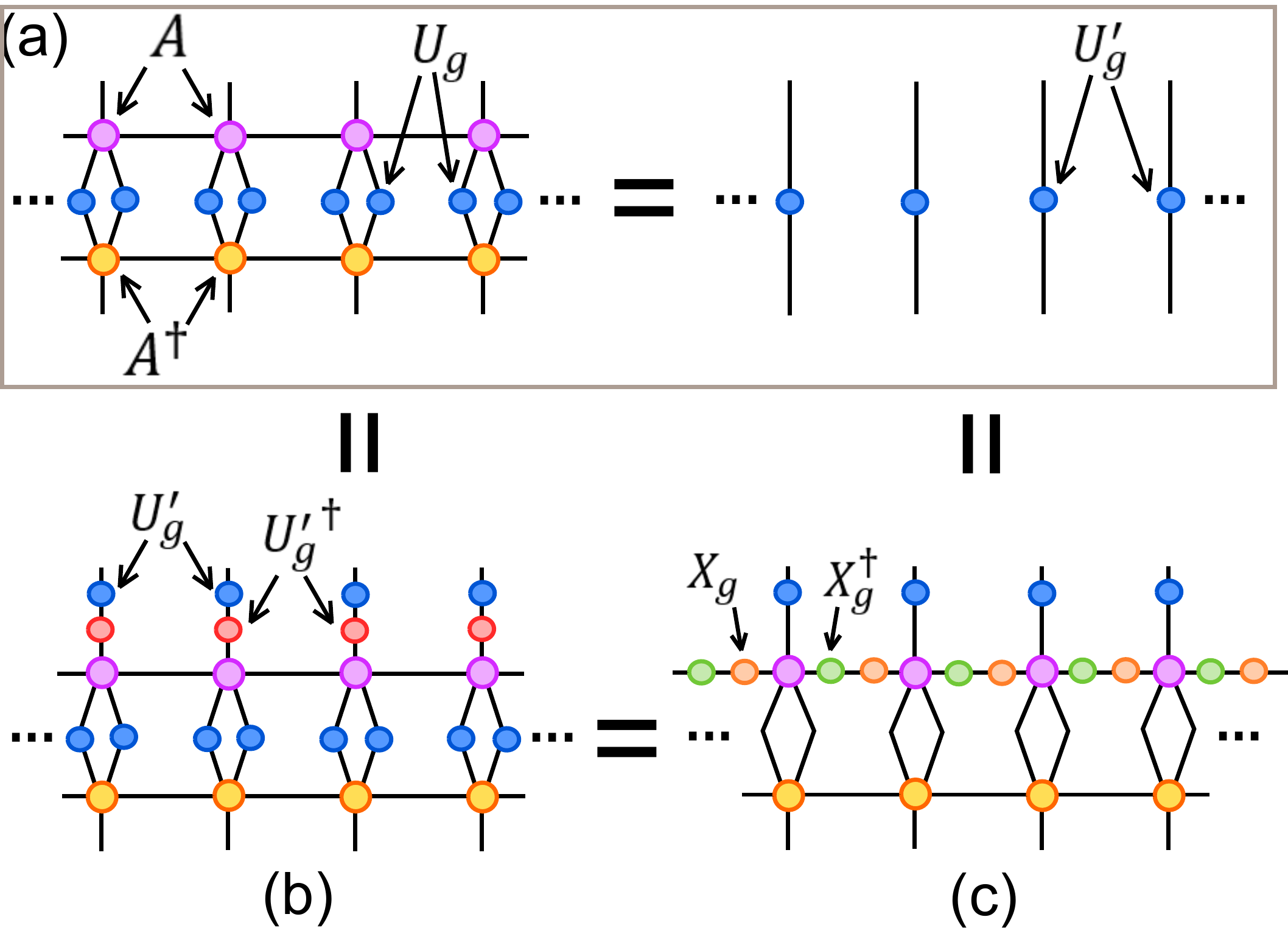}
\caption{\label{fig:symtensors} (a) Our main working assumption: a global on-site symmetry remains global and on-site after coarse-graining. This follows if tensor $A$ is symmetric [\eref{eq:merasymconstraint}] by the sequence of equalities (a) = (b) = (c) = (d). In (a) we introduced a resolution of identity $U'_g {U'_g}^\dagger$ on the top open indices. In (b) we applied \eref{eq:merasymconstraint}. In (c) we used $X_gX^\dagger_g = I$ and the fact that $u,w$ are isometries, \fref{fig:ER}(d).}
\end{figure}

We are now ready to address the following question: If a MERA describes a state with a global onsite symmetry $\mathcal{G}$ then is tensor $A$ (in each strip) symmetric? To proceed, consider the symmetry operators $O'_g$ on the coarse-grained lattice $\lattice'$, given by
\begin{equation} \label{eq:remainonsite}
O'_g \equiv W^\dagger O_g W, ~~\forall g \in \mathcal{G}.
\end{equation}
Note that without suitably constraining the individual $A$ tensors, or the transformation $W$ as a whole, the coarse-grained operators $O'_g$ are not necessarily on-site. However, in the rest of the paper, we will assume that the coarse-grained symmetry operators $O'_g$, in fact, remain on-site (see \fref{fig:symtensors}(a)) at all length scales. That is, the RG flow  \textit{preserves} the global and on-site character of the symmetry, at each step along the flow.

This assumption is consistent with the t'Hooft anomaly matching condition on the lattice, see e.g.  Ref.~\onlinecite{AnomaliesMERA}. A global symmetry that acts in an on-site way can be regarded as having a trivial t'Hooft anomaly, since such a global on-site symmetry can always be gauged. If anomalies are preserved along the RG flow, a trivial anomaly (that is, an on-site action) must remain trivial (on-site action) along the RG flow. In two dimensions, however, anomaly matching may be more subtle in the presence of topological order. Thus, our assumption is reasonable at least in one dimension.

Therefore, in this paper, we will regard the MERA description of a 1D symmetric ground state legitimate only if it is generated by an RG flow that preserves the global and on-site character of the symmetry.


Next, we observe that if tensor $A$ is symmetric \cite{symcontract} and transforms under the symmetry as
\begin{equation}\label{eq:merasymconstraint}
\vcenter{\hbox{\includegraphics[width=3.5cm]{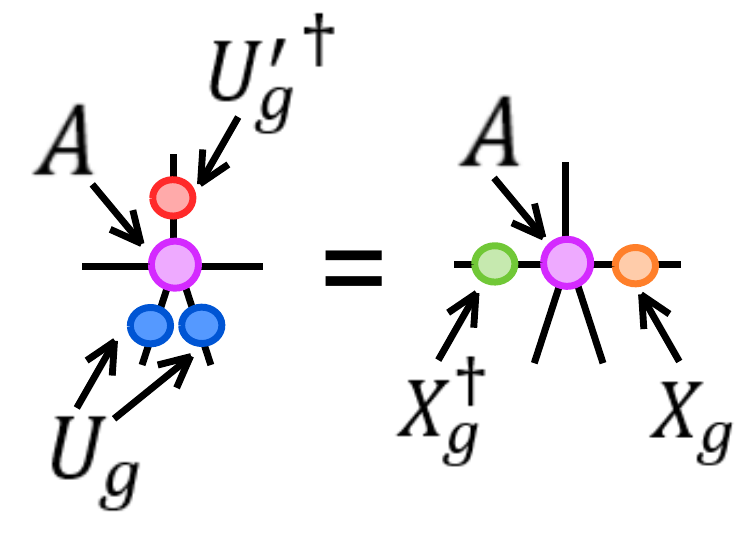}}},
\end{equation}
where $X_g$ is unitary, then the coarse-grained operators $O'_g$ are, in fact, on-site. This is demonstrated in  \fref{fig:symtensors}. Note that in \eref{eq:merasymconstraint} we will require that $U_g$ and $U'_g$ are linear representations of $\mathcal{G}$ but $X_g$ is allowed to be a projective representation of $\mathcal{G}$. ($U_g$ is required to be linear to allow for the possibility of symmetry fractionalizaton. Symmetric disentanglers and isometries then ensure that $U'_g,U''_g,\ldots$ are also linear.) \\
So we have the following implication:
\begin{equation}\label{eq:sofar}
\parbox{8em}{MERA made of \\ symmetric tensors} \Rightarrow \parbox{15em}{symmetry remains global \& \\ on-site under coarse-graining}
\end{equation}
However, it is not apparent that symmetric tensors are also \textit{necessary} for this implication. It is possible that a global on-site symmetry could generally be preserved by enforcing some \textit{global} constraints which are satisfied, say, by the  total tensor network without requiring the individual tensors to be symmetric.
But we show next that the implication (\ref{eq:sofar}), in fact, also holds in reverse, and therefore a global on-site symmetry can always be implemented in the MERA by means of symmetric tensors (assuming that the global and on-site character of the symmetry is preserved at all length scales).

\begin{figure}[t]
  \includegraphics[width=\columnwidth]{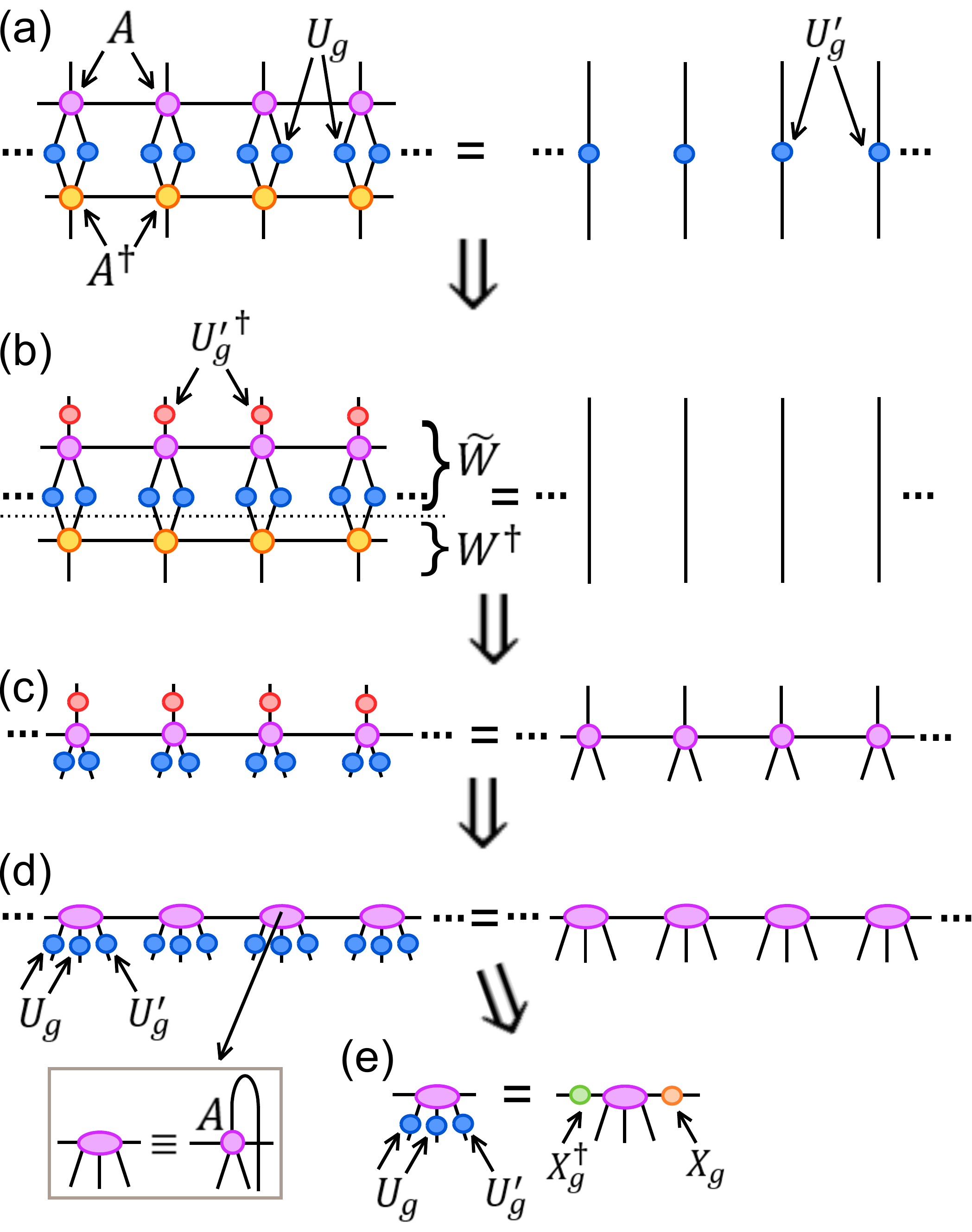}
\caption{\label{fig:layer} Proof of the reverse of implication (\ref{eq:sofar}). (a) We demand that on-site symmetry remains on-site under RG for all group elements $g \in G$. $U_g$ and $U'_g$ is the (linear) representation of the symmetry on each site of lattice $\mathcal{L}$ and $\mathcal{L}'$ respectively. (b) Re-organizing the previous equality by moving the symmetry operators on the right to the left hand side. (c) Since both $\tilde{W}$ and $W^\dagger$ are isometries, the equality shown in $(b)$ can only be true if $\tilde{W}$ is the adjoint of $W^\dagger$, that is, $\tilde{W} = W$ as depicted here. (d) $W$ expressed as a translation-invariant matrix product state $\ket{W}$ by bending the top indices. Equation depicted in \textit{(c)} translates to the MPS $\ket{W}$ having a global symmetry $\mathcal{G}$. (e) The local constraint fufilled by each MPS tensor as implied by the global on-site symmetry.}
\end{figure}

\subsection{Implication (\ref{eq:sofar}) also holds in reverse} 
To prove the reverse of (\ref{eq:sofar}), let us begin with our main working assumption, namely, the coarse-grained symmetry operators $O'_g$ remain on-site. This is depicted again in \fref{fig:layer}(a), and then re-arranged as shown in \fref{fig:layer}(b,c).
Next, we vectorize the MPO $W$ as a (translation-invariant) matrix product state (MPS) $\ket{W}$. Graphically, this corresponds to bending some indices of the MPO, see \fref{fig:layer}(d). Equation \fref{fig:layer}(c) implies that the MPS $\ket{W}$ has a global on-site symmetry $\mathcal{G}$, see \fref{fig:layer}(e). We could now recall \eref{eq:mpsSym} and conclude that tensor $A$ must be symmetric.
\begin{figure}[t]
  \includegraphics[width=7cm]{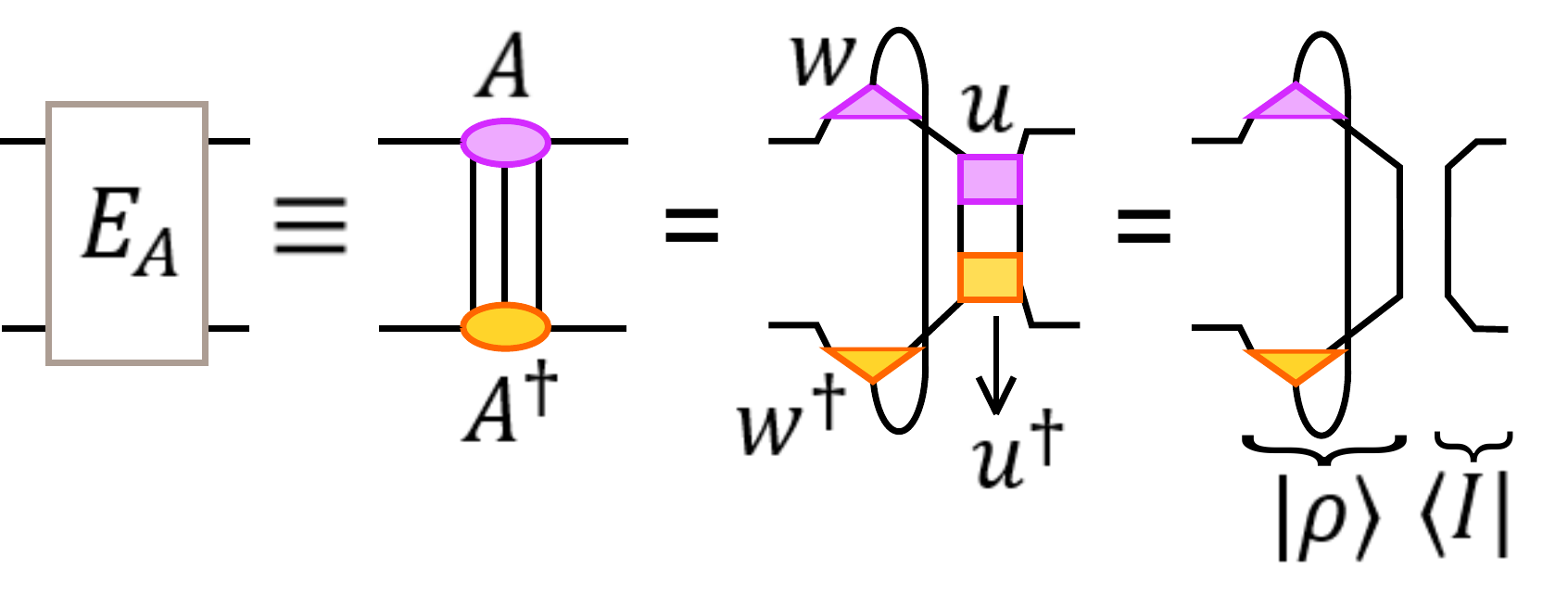}
\caption{\label{fig:transfer} (a) The matrix $E_A\equiv \sum_{i} A_i \otimes A^*_i$ is a rank-1 projector $\ket{\rho}\bra{I}$, since the isometry $u$ cancels out with its adjoint $u^\dagger$. Thus, MPS $\ket{W}$ is injective. $\ket{I}$ and $\ket{\rho}$ are the left and right eigenvectors of $E_A$ respectively. $\ket{I}$ is the (vectorized) identity matrix. $\ket{\rho}$ is obtained by contracting isometries $w$ and $w^\dagger$ as shown; this contraction can be viewed as acting with a completely positive map (whose kraus operators are given by $w$) on the identity, which results in a positive semidefinite matrix $\ket{\rho}$. Thus, MPS $\ket{W}$ is also in the canonical form.}
\end{figure}

However, \eref{eq:mpsSym} holds only for a symmetric MPS that is \textit{normal}. An MPS $B$ is said to be in the canonical form if the dominant eigenvalue of the matrix $E_B \equiv \sum_{i} B_i \otimes B^*_i$ is equal to 1, and the left (right) dominant eigenvector is the identity while the right (left) dominant eigenvector is a positive semi-definite matrix. If the dominant eigenvalue of $E_B$ is unique, then the MPS is also said to be normal.

MPS $\ket{W}$ is, in fact, both in the canonical form and normal. This is thanks to the isometric constraints \fref{fig:ER}(b), which are fulfilled by the MERA tensors, as shown in \fref{fig:transfer}. (MPS $\ket{W}$ is actually \textit{injective}, a property stronger than normality, see Appendix \ref{sec:appendixA}). 

Thus, applying \eref{eq:mpsSym} to the MPS $\ket{W}$, it follows that tensor $A$ must be symmetric as depicted in \fref{fig:layer}(e). By unvectorizing (bending back an index) \fref{fig:layer}(e), we obtain \eref{eq:merasymconstraint}. Thus, implication (\ref{eq:sofar}) also holds in reverse. We remark that the symmetric tensor $A$ can also, under reasonable assumptions, be decomposed into a symmetric disentangler $u$ and a symmetric isometry $w$, as shown in Appendix \ref{sec:appendixB}.

If the global on-site symmetry is preserved at all length scales, we can apply the above argument iteratively to all strips of the MERA. Thus, it follows that if a MERA has a global on-site symmetry and preserves the symmetry at all length scales, then its tensors must \textit{necessarily} be symmetric (up to gauge transformations). That is,
\begin{equation}\label{eq:sofarReverse}
\parbox{12.5em}{symmetry remains global \& \\ on-site under coarse-graining} \Rightarrow \parbox{8em}{MERA tensors \\ are symmetric}.
\end{equation}
Next, we turn to the question of symmetry fractionalization in the MERA.

\section{Gapped ground states}
In this section, we show that the symmetry \textit{does not} fractionalize in the MERA representation of a 1D \textit{gapped} symmetric ground state. Our strategy will be to translate an MPS description (possibly fractionalized) of a symmetry protected ground state to a MERA description, which we will examine for possible symmetry fractionalization. Recall, that the MERA describes the RG flow of a ground state. Therefore, to build the MERA description, we will coarse-grain the MPS (by means of entanglement renormalization) until a fixed point is reached.

\subsection{Symmetry does not fractionalize along the RG flow}  

First, let us consider a single coarse-graining step. Let MPS $B$ describe a ground state with a global on-site symmetry $\mathcal{G}$, and let MPS $B'$ denote the coarse-grained version obtained by means of entanglement renormalization. For convenience, we will assume that MPS $B$ satisfies \textit{Property 1} stated in Appendix \ref{sec:appendixA}, which also implies that it is normal and thus exhibits the local symmetry \eref{eq:mpsSym}. (1D symmetry protected ground states always admit an MPS description that satisfies \textit{Property 1}.\cite{mps3})

\begin{figure}[t]
  \includegraphics[width=6.5cm]{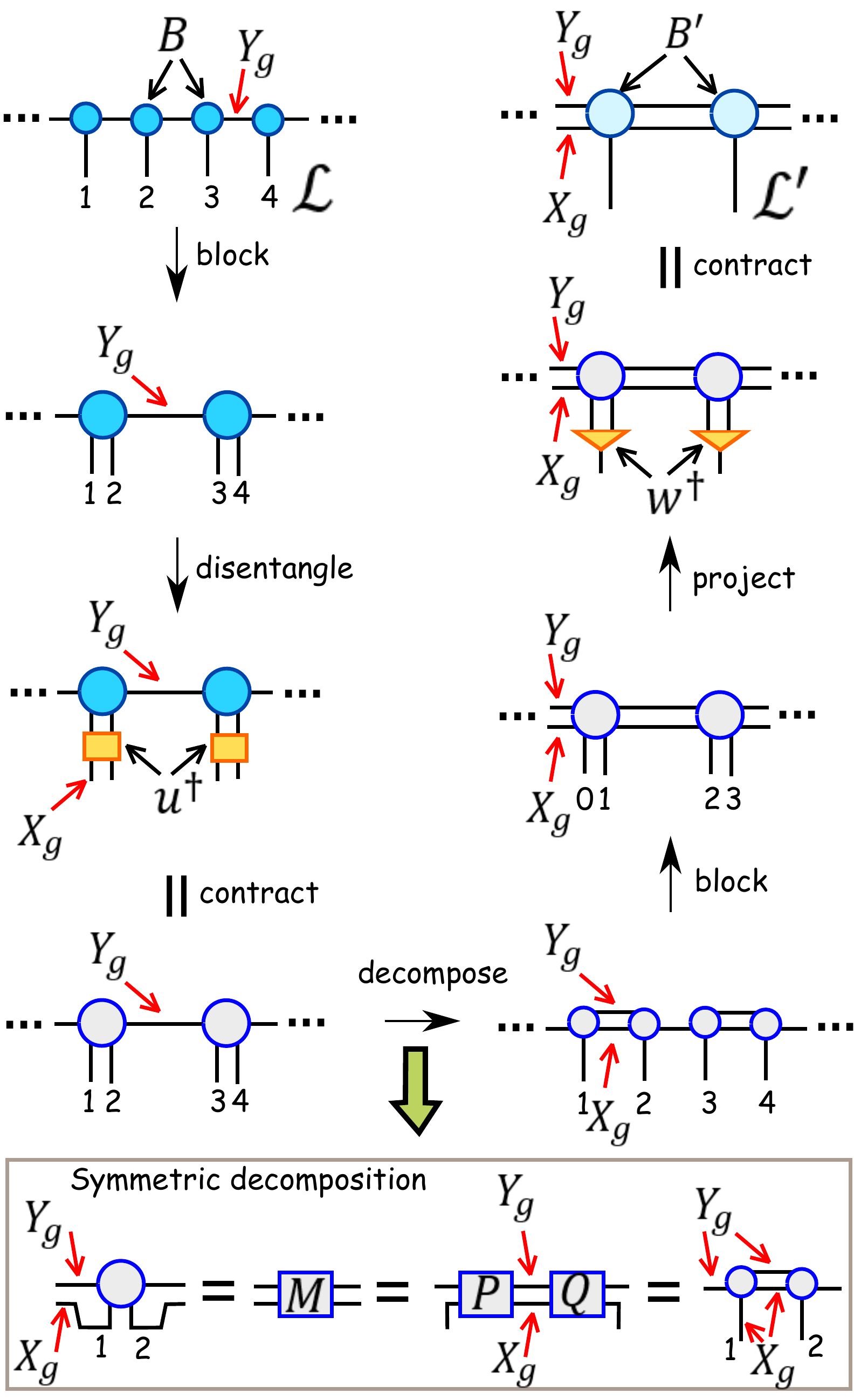}
\caption{\label{fig:update} Coarse-graining a normal MPS (top left) by means of entanglement renormalization, applied as a sequence of elementary operations. The resulting coarse-grained MPS is shown on the top right. Red arrows track the representations of the symmetry that appear on the bond indices of the intermediate MPS after each operation. The crucial decomposition step is elaborated in the box: A symmetric matrix $M$ can always be decomposed as a product of two symmetric matrices $P$ and $Q$, $M=PQ$. To see this, we note that $M$ is block diagonal in the symmetry basis (Schur's lemma). By applying a standard matrix decomposition such as eigenvalue decomposition to each block of $M$ separately, we can obtain factor matrices $P$ and $Q$, which are also block diagonal in the symmetry basis. Thus, $P$ and $Q$ are also symmetric.}
\end{figure}

The coarse-graining proceeds as the following sequence of elementary operations, see \fref{fig:update}: (i) block, for example, all odd pairs of sites and contract together the MPS tensors corresponding to each pair, (ii) apply on-site unitaries (the disentanglers) on the blocked MPS, (iii) decompose the resulting MPS back to the original lattice, (iv) block all even pairs of sites and contract together the MPS tensors corresponding to each pair, and finally (v) apply on-site isometries, which project to the support of the reduced density matrix of each blocked site.

We can determine how the resulting coarse-grained MPS tensor $B'$ transforms under the action of the symmetry, by tracking the action of the symmetry through this sequence of operations (as indicated in \fref{fig:update}). We find:
\begin{equation}\label{eq:coarsegrainedmps}
\vcenter{\hbox{\includegraphics[width=3.5cm]{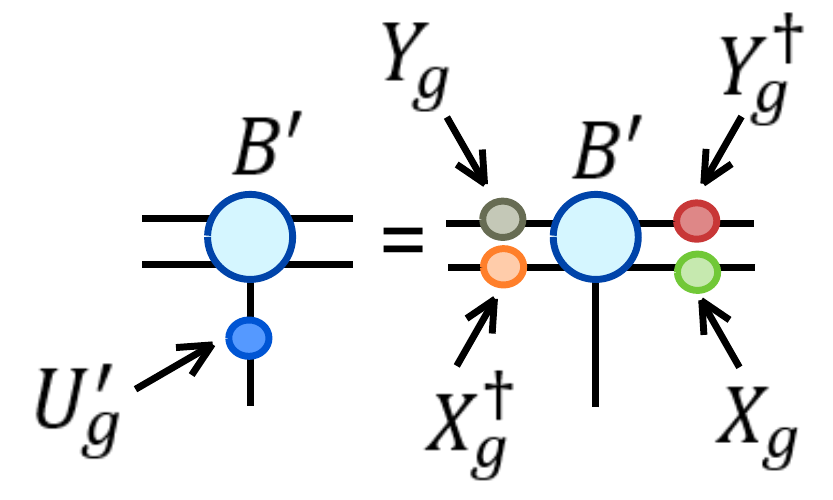}}}
\end{equation}
Notice that the representation $X_g$, which appears in the coarse-graining MPO via \eref{eq:merasymconstraint}, has been transferred to the coarse-grained MPS $B'$.

If the coarse-grained MPS $B'$ belongs to the same phase as MPS $B$, and is normal, it's bond representation $Y_g \otimes X^\dagger_g$ must belong to the same equivalence class as $Y_g$, that is, the bond representation of the input MPS $B$. Futhermore, this must be true if the input MPS $B$ belongs to \textit{any} symmetry protected phase, since the coarse-graining did not assume a specific phase to which MPS $B$ belongs. This is possible if, and only if, the representation $X_g$ is linear.\cite{projrep} Thus, the symmetry is not fractionalized in the coarse-graining MPO---which constitutes a strip of the MERA. 

Below we show that MPS $B'$ indeed satisfies the necessary criteria for the above argument: it belongs to the same phase as MPS $B$, and is normal. (Normality ensures that the bond representation can be trusted to correspond to the phase of the MPS.) To this end, we argue that each of the elementary operations (i)-(v), which implement the coarse-graining, preserve the symmetry, phase, and normality of the MPS.

\textit{Symmetry is preserved.---} Since the MPS tensors, the disentanglers, and isometries are all symmetric, the result of all the contraction steps is also a symmetric tensor \cite{symmetricTensors}, which ensures that the symmetry is preserved. The decomposition step, however, requires a more careful consideration. It turns out that a symmetric tensor can always be decomposed into symmetric tensors, if the decomposition is carried out blockwise, as explained in \fref{fig:update}.

\textit{Phase is preseved.---} Broadly speaking, two quantum many-body states are in the same phase if they can be connected by a finite depth circuit of finite range interactions; such states are  expected to have the same large length scale properties.\cite{CohmologyClassification,CohmologyClassification1,CohmologyClassification2} The disentanglers are local unitary transformations, and therefore keep the state in the phase by definition. The isometries are composed of local unitary transformations followed by a projection to the support of the local density matrix. Such a projection also preserves the ground state, and thus the phase of the state. (More generally, the disentanglers can also be isometries instead of unitaries, in which case they are also chosen to project to the support of the local reduced density matrices, an operation which preserves the phase.) The blocking and decomposition steps do not involve any truncation of the Hilbert space. Thus, they preserve all the information in the quantum state, including the phase it belongs to.

\textit{Normality is preserved.---} Follows from known properties of normal matrix product states, which are reviewed in Appendix \ref{sec:appendixA}.

Thanks to these properties, we can deduce that the representation $X_g$ is linear as argued previously.
Futhermore, iterating the above RG procedure, generates an RG flow where the resulting representation $X_g$ remains linear all along the RG flow.

\subsection{Symmetry also does not fractionalize at the RG fixed points}
The RG flow described in the previous section is also consistent with the expected RG fixed-point wavefunctions in a 1D symmetry protected phase. A representative RG fixed-point wavefunction $\ket{\Psi^{\mbox{\tiny fixed}}_\phi}$ in a symmetry protected phase that is labelled\cite{projrep} by $\phi \in H^2(\mathcal{G},U(1))$ is given by\cite{CohmologyClassification,CohmologyClassification1,CohmologyClassification2} :
\begin{equation}\label{eq:fixed}
\vcenter{\hbox{\includegraphics[width=7.5cm]{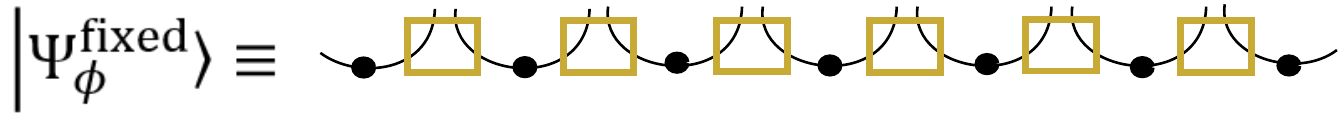}}}
\end{equation}
where $\ket{\Psi^-} \equiv \includegraphics[width=0.5cm]{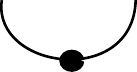}$ denotes a singlet under the action of a suitable \textit{projective} representation $V_{\phi}$ of $\mathcal{G}$, that is,
\begin{equation}
\ket{\Psi^-} = (V_{\phi} \otimes V_{\phi}) \ket{\Psi^-},
\end{equation}
and $\includegraphics[width=0.4cm]{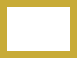}$ denotes a site of the lattice, which transforms as the \textit{linear} representation $(V_{\phi} \otimes V_{\phi})$. By `suitable' we mean $V_{\phi}$ is the smallest irreducible projective representation in the equivalence class labelled by $\phi \in H^2(\mathcal{G},U(1))$.

As described in Ref.~\onlinecite{symProtectedMERA}, the wavefunctions $\ket{\Psi^{\mbox{\tiny fixed}}_\phi}$  can be seen as fixed points of the coarse-graining transformation composed of the symmetric tensors:
\begin{equation}\label{eq:fixedtensors}
\vcenter{\hbox{\includegraphics[width=5.5cm]{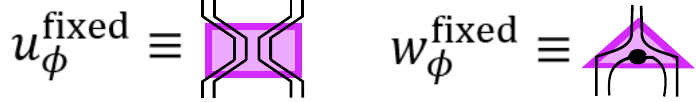}}}
\end{equation}
where $\includegraphics[width=0.15cm]{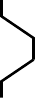}$ denotes the Identity in the irreducible representation $V_{\phi}$ and $\bra{\Psi^-} \equiv \includegraphics[width=0.5cm]{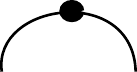}$. Notice that each index of both the tensors here correspond to double lines, and thus carries the \textit{linear} representation $(V_{\phi} \otimes V_{\phi})$.

However, as also described in Ref.~\onlinecite{symProtectedMERA}, if symmetry fractionalization is allowed then the wavefunctions $\{\ket{\Psi^{\mbox{\tiny fixed}}_\phi}\}_\phi$ are no longer fixed points, since they can all be coarse-grained to a product state by means of the tensors:
\begin{equation}\label{eq:tensorsphi}
\vcenter{\hbox{\includegraphics[width=4.5cm]{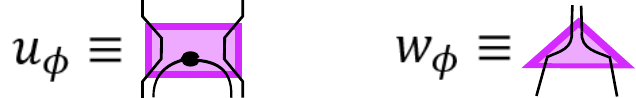}}}
\end{equation}
This means that entanglement renormalization reproduces the expected fixed point $\ket{\Psi^{\mbox{\tiny fixed}}_\phi}$ in each symmetry protected phase, \textit{provided} the symmetry does not fractionalize. We have shown that this is in fact the case, and therefore also confirm the observations presented in Ref.~\onlinecite{symProtectedMERA}.

The MERA description of a ground state belonging to phase $\phi$ consists of the RG flow to the fixed-point state $\ket{\Psi^{\mbox{\tiny fixed}}_\phi}$. We have seen that the symmetry does not fractionalize during the RG flow or at the fixed point. Thus we conclude that the symmetry \textit{does not} fractionalize in the MERA description of a \textit{gapped} symmetry protected ground state.

We remark that symmetry fractionalization in the MPS has been exploited to devise practical schemes to detect symmetry protected phases in MPS ground state simulations: either by computing non-local order parameters\cite{phaseDetection1} (topological invariants) or by making use of symmetric tensors to directly detect the equivalence class of the bond representation for the state\cite{symMPSDetection}. In the MERA, one must instead examine the fixed point of the RG flow to detect the phase, see Refs.~\onlinecite{symProtectedMERA, symProtectedMERA1}.

\section{Critical ground states}
In the previous section, we argued that the symmetry does not fractionalize in a MERA description of a \textit{gapped} symmetry protected ground state. The argument relied on the fact that a gapped ground state admits a faithful MPS description. On the other hand, a \textit{critical} ground state cannot be faithfully described as a MPS; therefore, our argument does not carry over to critical states.

However, we expect that the RG flow at a critical point also preserves the global and on-site character of the symmetry---thus, the local symmetry constraint \eref{eq:merasymconstraint} still holds. This means that the MERA description of a critical ground is also composed of symmetric tensors (up to gauge transformations), \eref{eq:sofarReverse}. Therefore, it is at least possible that the symmetry can fractionalize in the MERA description of a \textit{critical} symmetry protected ground state. (Since symmetric tensors were a prerequisite for symmetry fractionalization in the MPS.)

A possible argument against symmetry fractionalization in critical systems could be that perhaps symmetry fractionalization results from, and is intimately tied to, the short-range entanglement structure that is characteristic of gapped ground states. However, below we illustrate a counter-example.

Consider the MERA composed from copies of the following tensors:
\begin{equation}\label{eq:criticaltensors}
\vcenter{\hbox{\includegraphics[width=2.5cm]{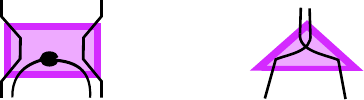}}}
\end{equation}
These are almost the same tensors as the ones in \eref{eq:tensorsphi}, except that we have swapped the top indices in the $w_\phi$ tensor. The symmetry is clearly still fractionalized on some of the indices.

\begin{figure}[t]
  \includegraphics[width=\columnwidth]{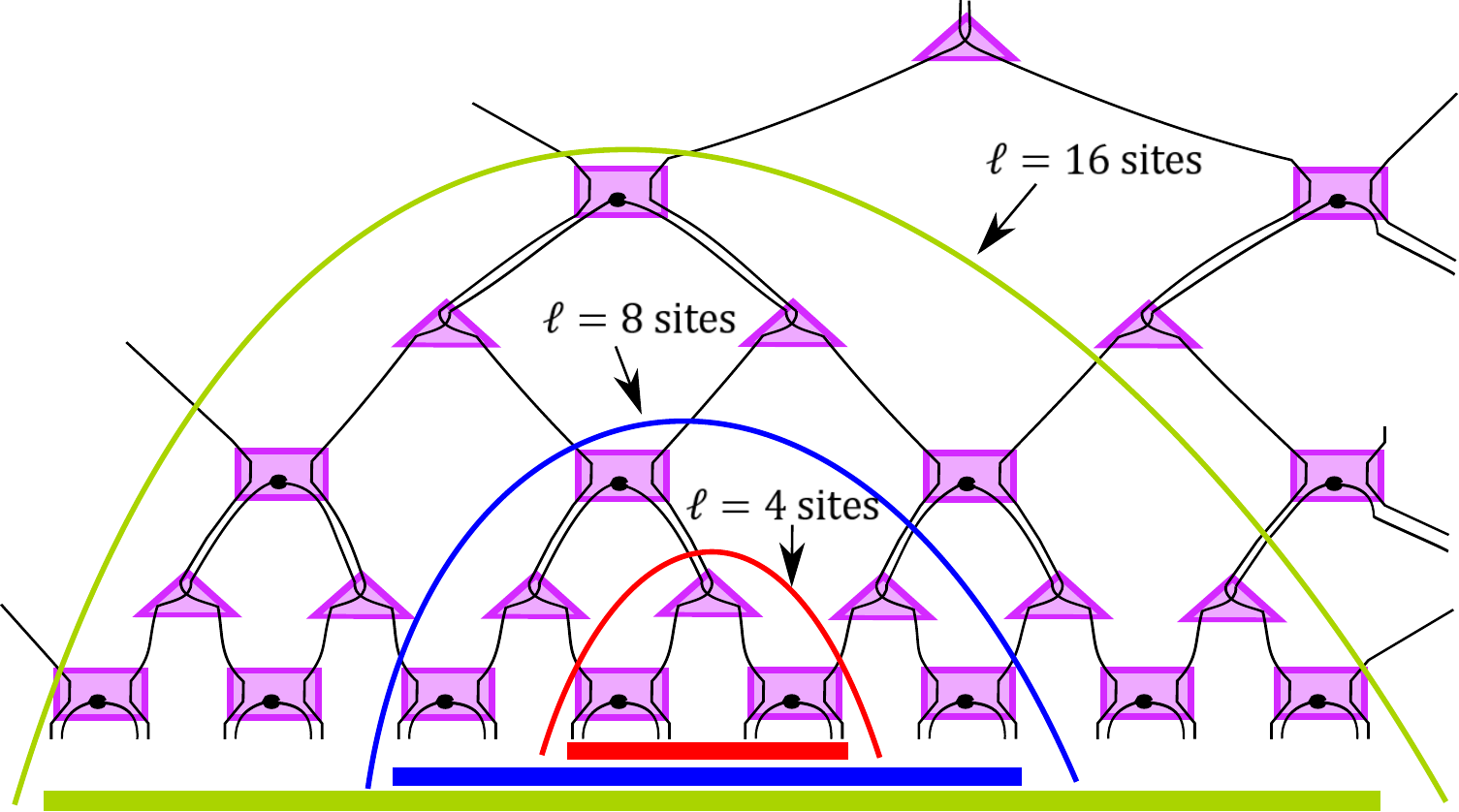}
\caption{\label{fig:criticalSPT} A patch of the MERA composed of copies of tensors shown in \eref{eq:criticaltensors}. This MERA represents a state that exhibits both symmetry fractionalization and long-range entanglement. The latter can be deduced by considering blocks of sites (open indices) on the 1D lattice with increasing number of sites e.g. illustrated here are blocks of $\ell=4,8$ and $16$ sites indicated in red, blue, and green respectively. The entanglement of each block with the rest of the lattice is proportional to the number of singlets shared between the block and the remaining lattice; these are the singlets that are intersected by the geodesic (shown in the respective colors) that extends between the end-points of the block through the tensor network. For the three blocks illustrated here the respective geodesic intersects $4,6$ and $8$ sites. Generally, a block of $\ell$ sites shares approximately log $\ell$ singlets with the remaining lattice.}
\end{figure}

However, in contrast to the MERA from \eref{eq:tensorsphi}, the average entanglement entropy of a block of $\ell$ sites in the MERA from \eref{eq:criticaltensors} grows as log $\ell$, as illustrated in \fref{fig:criticalSPT}. This is also the characteristic scaling of entanglement entropy in 1D critical ground states. Note, however, that the state represented by this MERA (\eref{eq:criticaltensors}) cannot be the ground state of a critical system, since, for example, it is not translation-invariant. One can speculate that it may be possible to construct a fractionalized critical ground state by taking superpositions of translations of such a MERA. Nevertheless, this example illustrates that a MERA can exhibit both symmetry fractionalization and long range entanglement characteristic of 1D critical systems.

If the symmetry can fractionalize at a critical point, however, there are at least a couple of potentially interesting questions one can ask. First, analogous to the MPS, can symmetry fractionalization in the MERA be used to characterize \textit{critical} symmetry protected phases which were, for example, introduced recently in Refs.~\onlinecite{symProtectedCritical,symProtectedCritical1,AnomaliesMERA}. Second, how does symmetry fractionalization in the MERA interplay with the emergent conformal symmetry at the critical fixed point. And third, what role, if any, does symmetry fractionalization play in holographic interpretations of the MERA.

\section{Summary} 
We first showed that a MERA with a global on-site symmetry necessarily consists of symmetric tensors (up to gauge transformations), provided we assume that the symmetry remains global and on-site along the RG flow. This motivates the existing use of symmetric tensors for implementing global on-site symmetries in MERA algorithms, and also sets the stage for exploring symmetry fractionalization in the MERA. Subsequently, we argued that the symmetry does not fractionalize in MERA descriptions of 1D gapped symmetry protected ground states. However, without imposing any other constraints on the RG flow, symmetry fractionalization can still occur in MERA descriptions of 1D critical symmetry protected ground states, which could potentially lead to interesting applications. 

\textbf{Acknowledgements.---} SS thanks Frank Verstraete for useful discussions and acknowledges the support provided by the Alexander von Humboldt Foundation and the Federal Ministry for Education and Research through the Sofja Kovalevskaja Award. This research was funded in part by the Australian Research
Council Centre of Excellence for Engineered Quantum
Systems (Project number CE170100009)


\appendix
\section{Normal Matrix Product States}\label{sec:appendixA}
In this appendix, we briefly review the definition and some standard properties of \textit{normal matrix product states}, which we used in this paper. The following material has been taken mostly from Refs.~\onlinecite{MPSSym,MPSSym2,mps3}.

A translation-invariant MPS $\ket{\Psi}$ of a lattice $\mathcal{L}$ with $L$ sites is composed from a set of matrices $\{A_i\}_{i=1}^{d}$ as
\begin{equation}\label{eq:MPS}
\ket{\Psi} = \sum_{i_1i_2\ldots i_L} \mbox{Tr}(A_{i_1}A_{i_2}\ldots A_{i_L}) \ket{i_1} \otimes \ket{i_2}  \ldots \otimes \ket{i_L},
\end{equation}
where $i_k$ labels an orthonormal basis on site $k$. The matrices $\{A_i\}$ can be glued together to obtain a 3-index tensor $A$, such that fixing a particular value of the index $i$ selects a matrix $A_i$ from the tensor. We succintly refer to $\ket{\Psi}$ as MPS $A$. We will also represent the MPS graphically as shown below:
\begin{equation}
\vcenter{\hbox{\includegraphics[width=5cm]{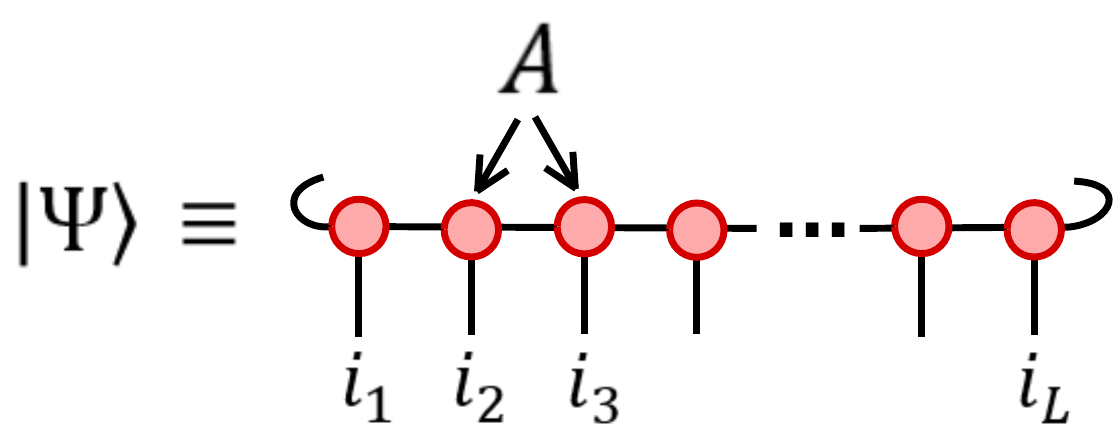}}} \nonumber
\end{equation}
Each circle represents a copy of tensor $A$. Each open index $i_k$ labels an orthonormal basis $\ket{i_k}$ on site $k$ of $\mathcal{L}$. The probability amplitude for a given configuration $\ket{i_1} \otimes \ket{i_2}  \ldots \otimes \ket{i_L}$ is obtained by fixing the open indices to the corresponding values and performing the trace of the product of the resulting selection of matrices.\\
\textbf{Definition 1. (Injectivity)} Let us define a matrix $M_A$ as 
\begin{equation}\label{eq:MA}
\vcenter{\hbox{\includegraphics[width=2.5cm]{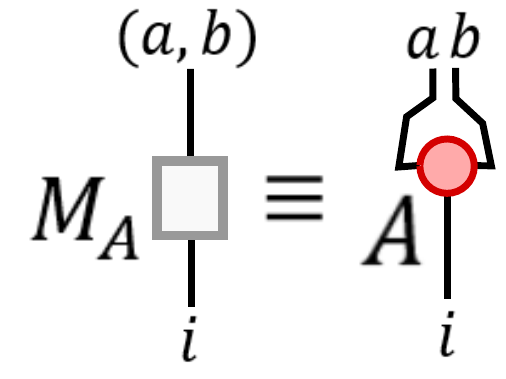}}}
\end{equation}
whose rows are indexed by the pair $(a,b)$ and columns are indexed by $i$. MPS $A$ is said to be \textit{injective} if the matrix $M_A$ has a (pseudo-)inverse $M^{-1}_A$, $M_A M^{-1}_A = I$. This can also be expressed directly in terms of $A$ as: there exists a 3-index tensor $A^{-1}$ such that
\begin{equation}
\vcenter{\hbox{\includegraphics[width=2cm]{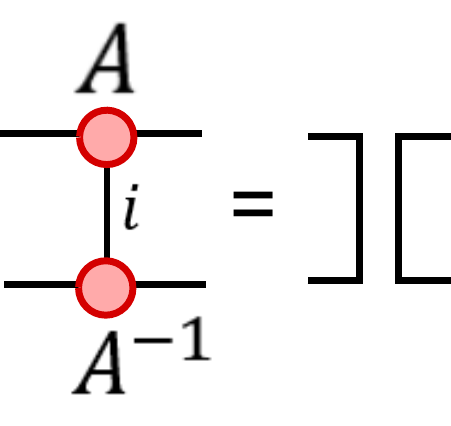}}}
\end{equation}
\textbf{Definition 2. (Normality)} Consider another lattice $\mathcal{L}^{\times s}$ that is obtained by blocking together $s$ sites of $\mathcal{L}$. State $\ket{\Psi}$ can be described on the lattice $\mathcal{L}^{\times s}$ by an MPS whose matrices $\{A^{\times s}_j\}$ are given by
\begin{equation}
A^{\times s}_j \equiv A_{i_k} A_{i_{k+1}}\ldots A_{i_{k+s}},
\end{equation}  
$j \equiv (i_k,i_{k+1},\ldots,i_{k+s})$ labels an orthonormal basis $\ket{j} \equiv \ket{i_k} \otimes \ket{i_{k+1}} \ldots \otimes \ket{i_{k+s}}$ on a site of $\mathcal{L}^{\times s}$. For example, for $s=2$ we have
\begin{equation}\label{eq:block2}
\vcenter{\hbox{\includegraphics[width=3cm]{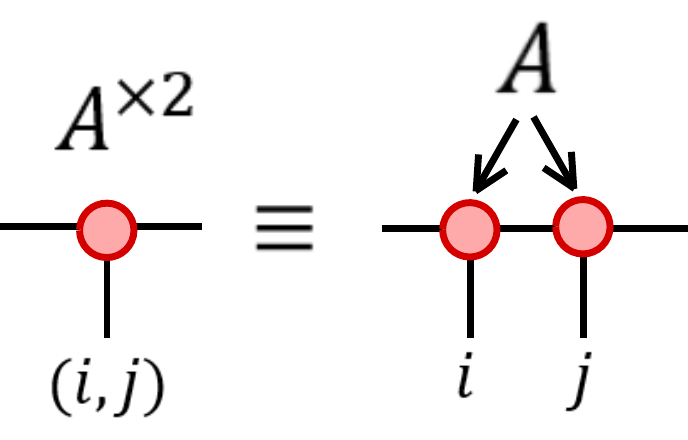}}}
\end{equation}
MPS $A$ is called \textit{normal} if there exists an $s$ such that the MPS $A^{\times s}$ is injective.\\
\textbf{Definition 3. (Canonical form)} Define the map $E_A(\circ) = \sum_{i} A_{i} (\circ) A^\dagger_{i}$, depicted as
\begin{equation}\label{eq:transfer}
\vcenter{\hbox{\includegraphics[width=2cm]{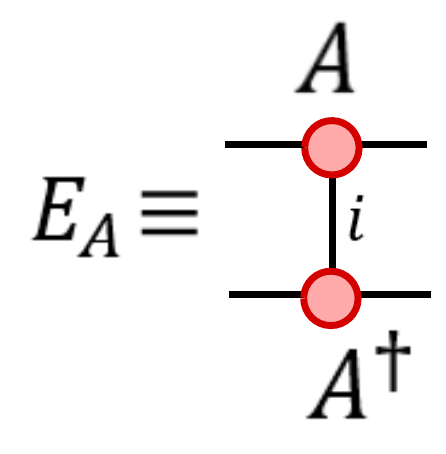}}}
\end{equation}
where $\dagger$ denotes the Hermitian adjoint. MPS $A$ is said to be in the \textit{canonical form} if (i) the spectral radius of the map $E_A$ is equal to one, and (ii) $E_A$ has an eigenvalue equal to its spectral radius (i.e. =1) and (iii) the left (right) dominant eigenvector is the identity while the right (left) dominant eigenvector is a positive semi-definite matrix.\\
\textbf{Property 1. (A sufficient condition for normality.)} If MPS $A$ (i) is in the canonical form and (ii) the largest eigenvalue of $E_A(X)$ is equal to 1 and non-degenerate, then it is normal.\\
See e.g. Proposition II.1. in Ref.~\onlinecite{MPSSym2}.\\
In the remainder, we will assume that MPS $A$ satisfies \textit{Property 1} (and is thus normal).\\
\textbf{Property 2. (Blocking)} MPS $A^{\times s}$ is also normal.\\
\textit{Proof.} Since $E_{A^{\times s}} = (E_A)^s$, the spectral properties of the map $E_{A^{\times s}}$ satisfy the conditions for normality demanded in \textit{Property 1}.\\
\textbf{Property 3. (Equivalence)} If a normal MPS $A$ is equal (up to a phase) to another normal MPS $A'$ then there exists an invertible matrix $X$ such that $A'_i = XA_iX^{-1}, \forall i$, depicted as
\begin{equation} \label{eq:equivalentNormalMPS}
\vcenter{\hbox{\includegraphics[width=3.5cm]{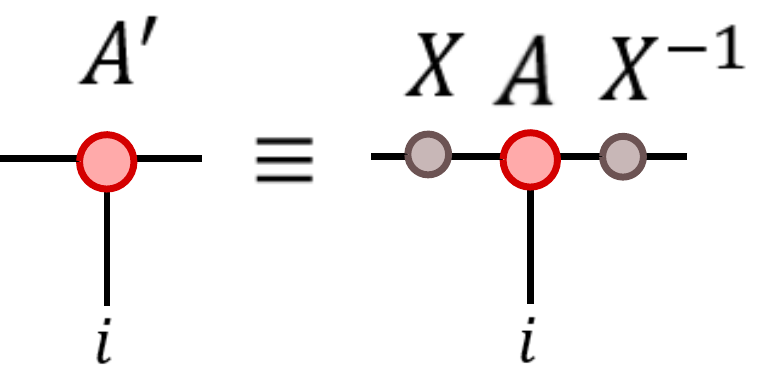}}}
\end{equation}
Proved as Theorem 7 in Ref.~\onlinecite{mps3}.\\
\textbf{Property 4. (Decomposition)} Consider MPS $A^{\times 2}$ obtained from MPS $A$ by blocking two sites, \eref{eq:block2}. Let us perform an arbitrary decomposition $A^{\times 2}_{ij} \equiv P_iQ_j$ and obtain a new MPS ${A'}^{\times 2}_{ij} \equiv Q_iP_j$,
\begin{equation}
\vcenter{\hbox{\includegraphics[width=7cm]{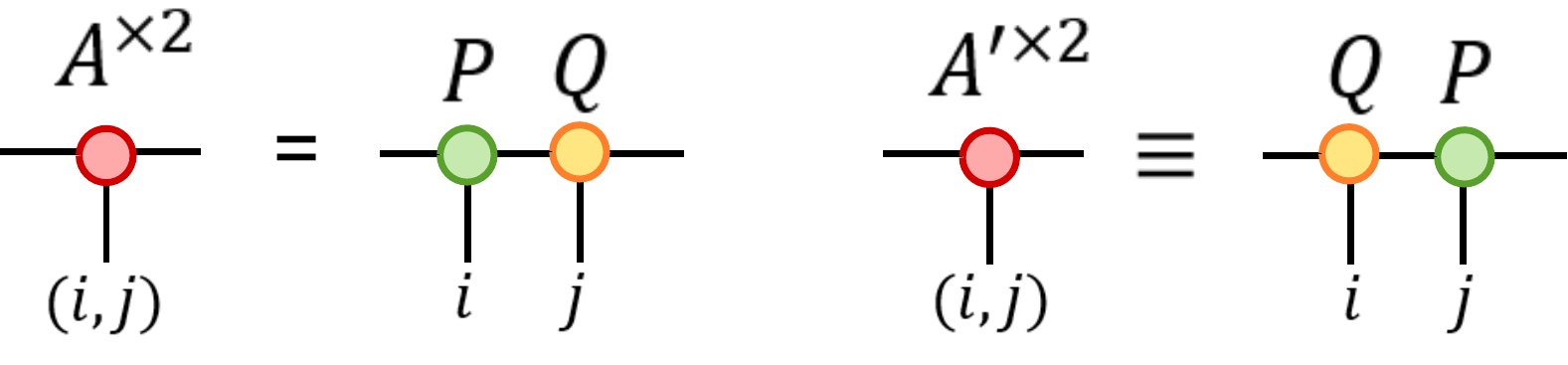}}}.
\end{equation}
MPS ${A'}^{\times 2}$ is also normal.\\
\textit{Proof.} Clearly, MPS $A^{\times 2}$ and MPS ${A'}^{\times 2}$ both represent the same state. (Since the trace in \eref{eq:MPS} remains unaffected by the decomposition considered here.)  Thus, the two MPSs are equivalent and related according to \eref{eq:equivalentNormalMPS}, that is, there exists an invertible matrix $Y$ such that
\begin{equation}
\vcenter{\hbox{\includegraphics[width=3.5cm]{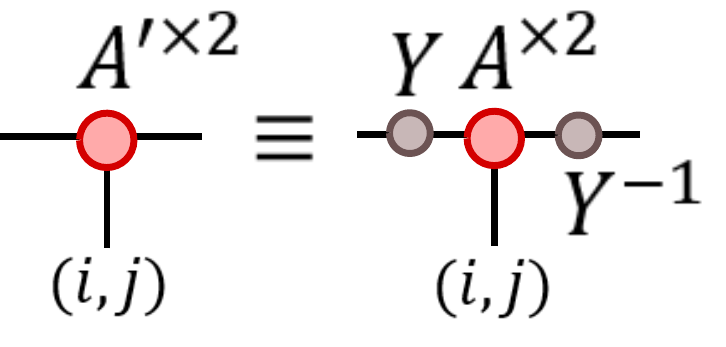}}}
\end{equation}
As a result, the map $E_{{A'}^{\times 2}}$ is related to the map $E_{{A}^{\times 2}}$ as
\begin{equation}
\vcenter{\hbox{\includegraphics[width=4.25cm]{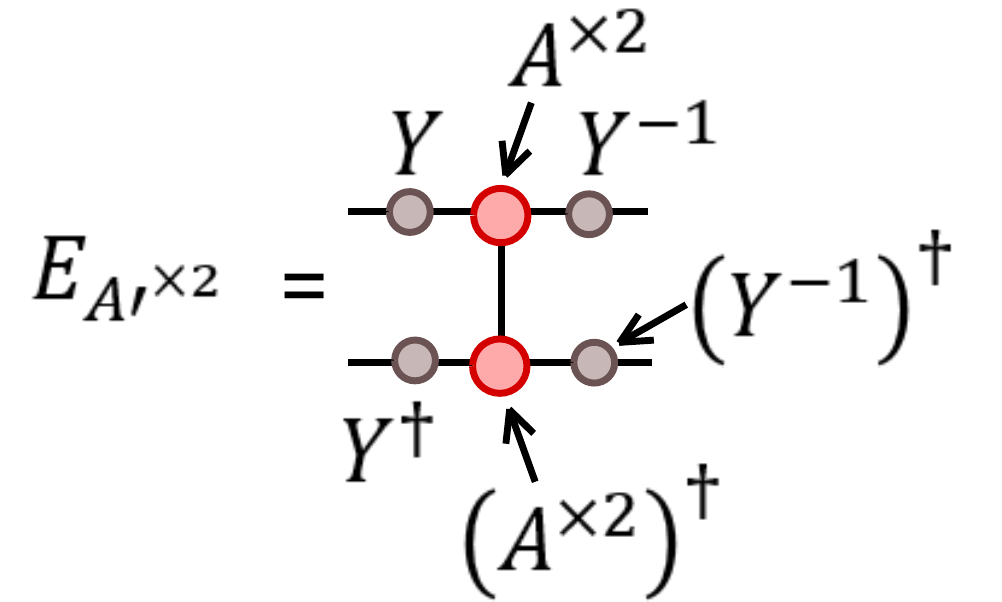}}}
\end{equation}
Thus, both maps have the same spectrum. We know that the dominant left and right eigenvectors of $E_{{A}^{\times 2}}$ are positive-semidefinite (in fact, one of the eigenvectors is the identity). Since the left and right eigenvectors of $E_{{A'}^{\times 2}}$ are related to those of $E_{{A}^{\times 2}}$ by conjugation under $Y(.)Y^\dagger$ [$Y^{-1}(.)(Y^{-1})^\dagger$], the dominant left and right eigenvectors of $E_{{A'}^{\times 2}}$ are also positive-semidefinite (since positive-definiteness is preserved under any conjugation of the form $Z(.)Z^\dagger$).\\
\textbf{Property 5. (On-site unitary)} Consider MPS $A'$ obtained from MPS $A$ by acting with unitary $U$ on each site of $\mathcal{L}$. We have $A'_i = \sum_j U_{ij}A_j$, depicted as
\begin{equation}
\vcenter{\hbox{\includegraphics[width=2.5cm]{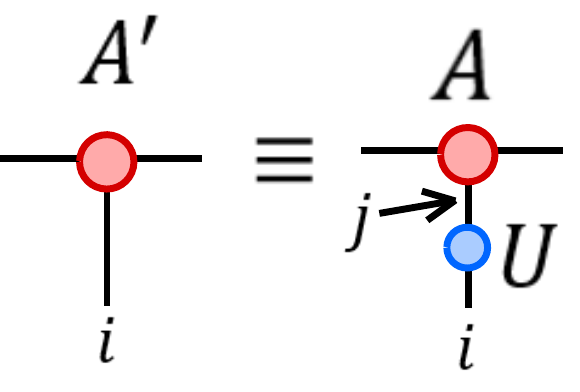}}}.
\end{equation}
MPS $A'$ is also normal.\\
\textit{Proof.} Follows from the fact that the maps $E_A$ and $E_{A'}$ [\eref{eq:transfer})] are equal.\\
\textbf{Property 6.} Let $M_A = usv$ denote the singular value decomposition of the matrix $M_A$ of \eref{eq:MA} where $u^\dagger u = \mathbb{I},v v^\dagger = \mathbb{I},$ and $s$ is a diagonal matrix with positive entries (depicted below).
\begin{equation} \label{eq:A_svd}
\vcenter{\hbox{\includegraphics[width=2.75cm]{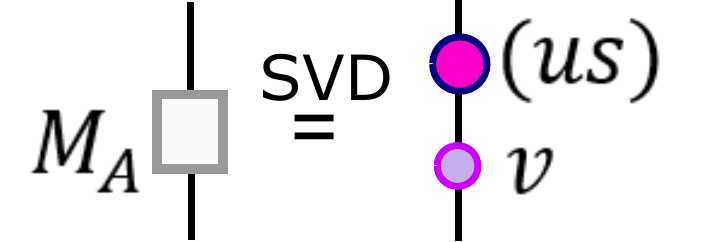}}}.
\end{equation}
Then the range of the matrix $v$ contains the support of the one-site density matrix obtained from MPS $A$.\\ 
\textit{Proof.} The one-site density matrix $\rho$ is given by
\begin{equation}
\vcenter{\hbox{\includegraphics[width=3.5cm]{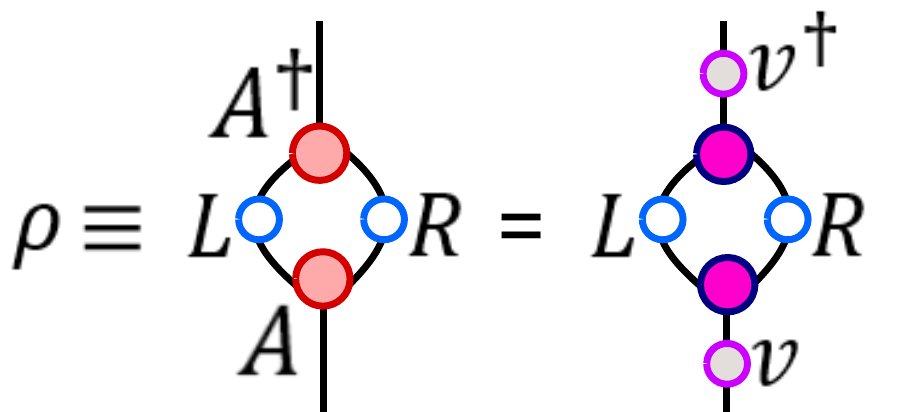}}},
\end{equation}
where $L,R$ are the dominant left and right eigenvectors of $E_A$ respectively. The equality is obtained by replacing tensor $A$ (and $A^{\dagger}$) by the singular value decomposition \eref{eq:A_svd}. Consequently, $v\rho v^\dagger$ has the same spectrum as $\rho$ (since $v v^\dagger = I$). Thus, $v$ preserves the support of $\rho$.\\  
\textbf{Property 7. (On-site projector)} Consider MPS $A'$ obtained from MPS $A$ by acting with the isometry $v$ of \eref{eq:A_svd} on each site of $\mathcal{L}$. (The isometry $v$ projects to the support of the one-site density matrix, Property 6). We have $A'_i = \sum_j v_{ij} A_j$, depicted below
\begin{equation}
\vcenter{\hbox{\includegraphics[width=2.5cm]{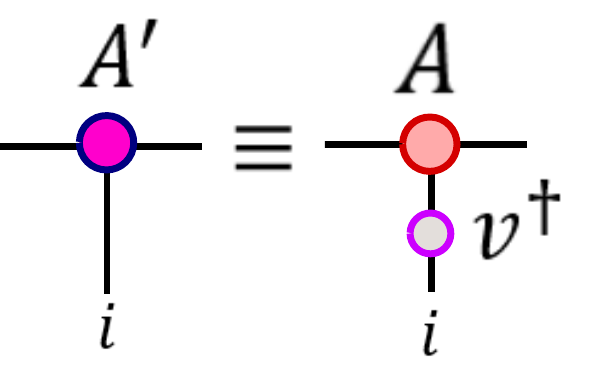}}}.
\end{equation}
MPS $A'$ is also normal.\\
\textit{Proof.} Follows from the fact that the maps $E_A$ and $E_{A'}$ [\eref{eq:transfer}] are equal, since
\begin{equation}
\vcenter{\hbox{\includegraphics[width=5.25cm]{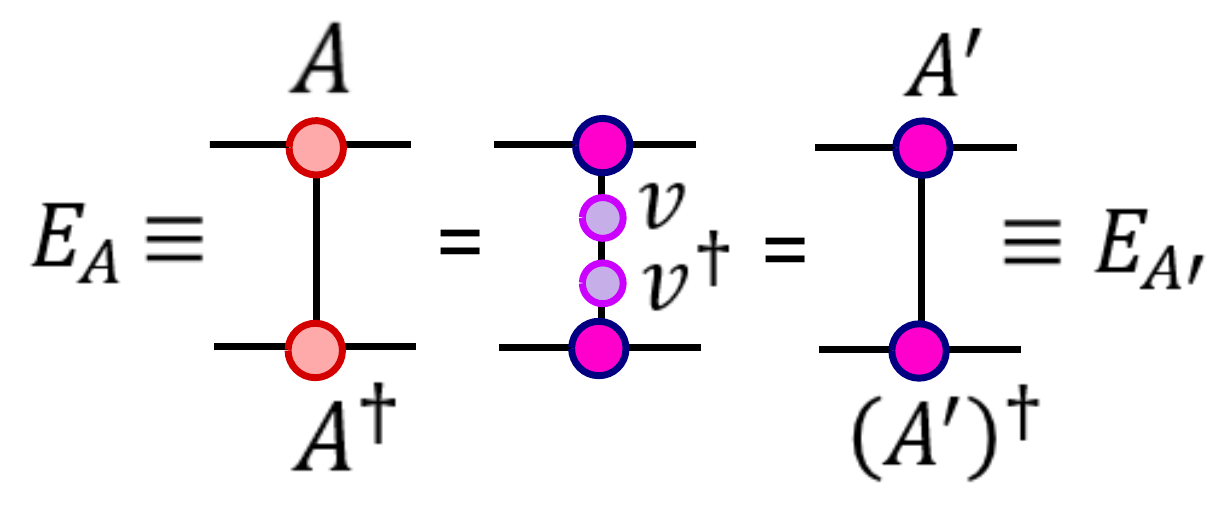}}}
\end{equation}
where we used $v v^\dagger = \mathbb{I}$.

\section{Symmetric disentanglers and isometries from tensor $A$}\label{sec:appendixB}

Consider a symmetric tensor $A$ that fulfills \eref{eq:merasymconstraint}, which we recall below:
\begin{equation}\label{eq:B1}
\vcenter{\hbox{\includegraphics[width=3.5cm]{merasymconstraint}}}
\end{equation}
Recall that tensor $A$ is obtained by contracting a disentangler $u$ and an isometry $w$ as
\begin{equation}\label{eq:B2}
\vcenter{\hbox{\includegraphics[width=3cm]{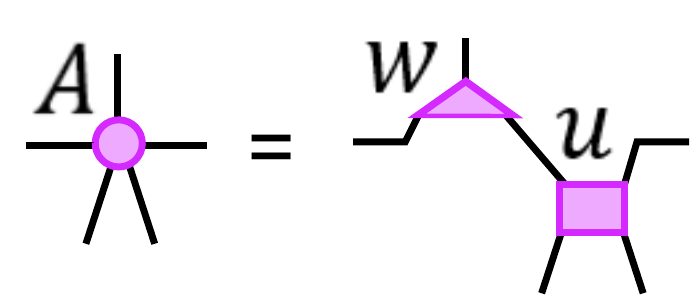}}}
\end{equation}
Below we show tensors $u$ and $w$ can always be chosen to be symmetric (under reasonable assumptions). \\
Let us contract a disentanger and an isometry differently to obtain another tensor $A'$ as
\begin{equation}
\vcenter{\hbox{\includegraphics[width=3cm]{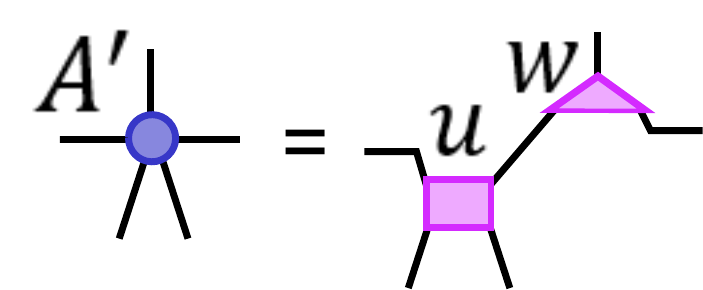}}}
\end{equation}
where, in comparison to \eref{eq:B2}, we have shifted the relative positions of $u$ and $w$. We can now repeat the argument in the main text to prove that tensor $A'$ is also symmetric. (Since the argument did not depend on how the MERA tensors were paired and combined to form a matrix product operator.) Thus, we also have
\begin{equation}
\vcenter{\hbox{\includegraphics[width=3.5cm]{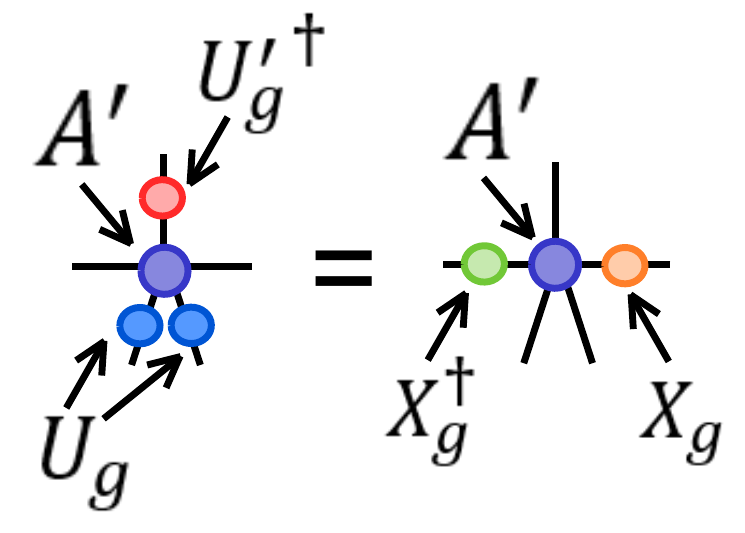}}}
\end{equation}
Next, define tensor $Q$ obtained by contracting $A'$ and $A^\dagger$ as
\begin{equation}
\vcenter{\hbox{\includegraphics[width=6.5cm]{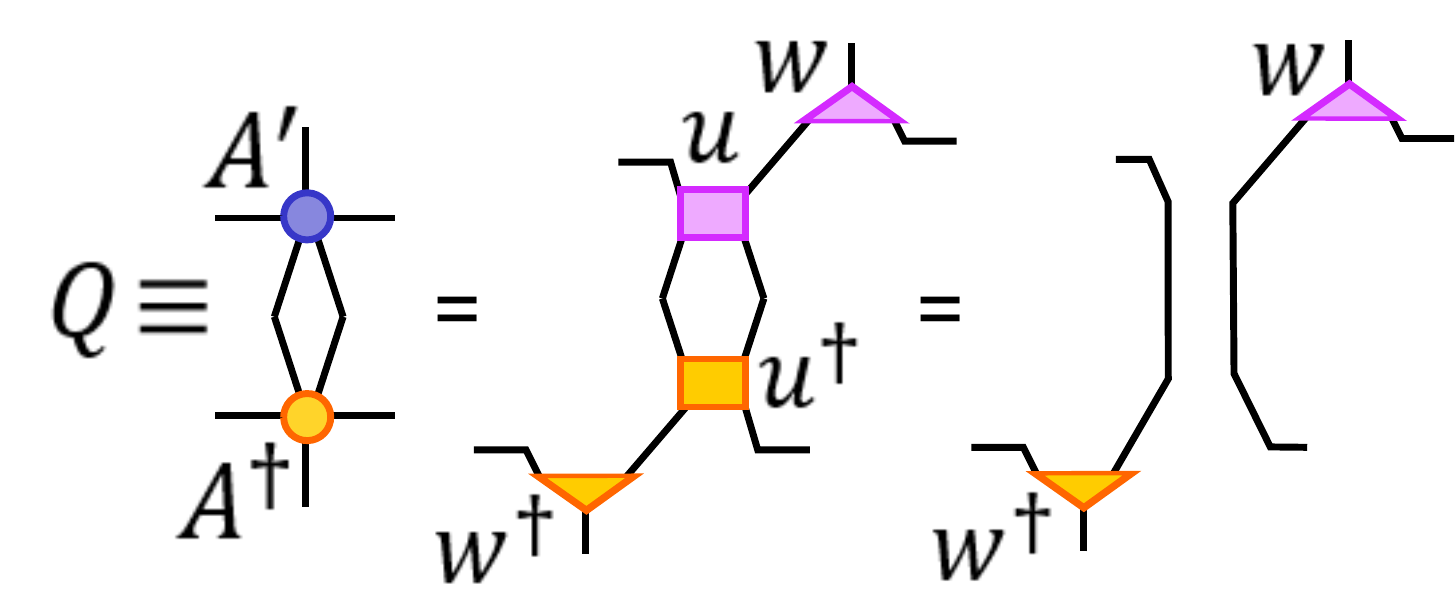}}}
\end{equation}
Tensor $Q$ is symmetric since it was obtained by contracting symmetric tensors.\cite{symmetricTensors} Since $Q = w^\dagger \otimes w$, $w$ must be symmetric fulfilling:
\begin{equation}
\vcenter{\hbox{\includegraphics[width=2.5cm]{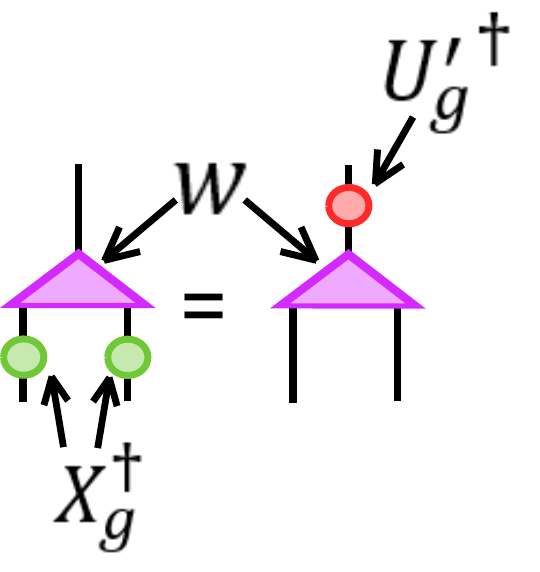}}}
\end{equation}
In order to show that $u$ is also symmetric, we have to assume $w$ has a pseudo-inverse $w^{-1}$ (as shown below)
\begin{equation}
\vcenter{\hbox{\includegraphics[width=2.5cm]{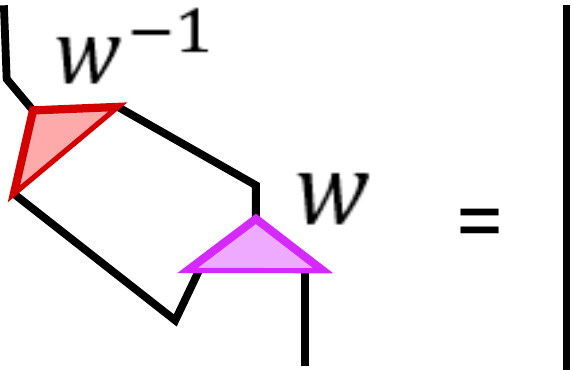}}}
\end{equation}
It can be easily shown that if a pseudo-inverse exists, it must be symmetric. (Tensor $w$ is block-diagonal when expressed in the symmetry basis for each index. A symmetric pseudo-inverse of $w$ is obtained by replacing each block with its pseudo-inverse.) If $w^{-1}$ exists then we can act with it on both sides of \eref{eq:B2} and obtain
\begin{equation}
\vcenter{\hbox{\includegraphics[width=5cm]{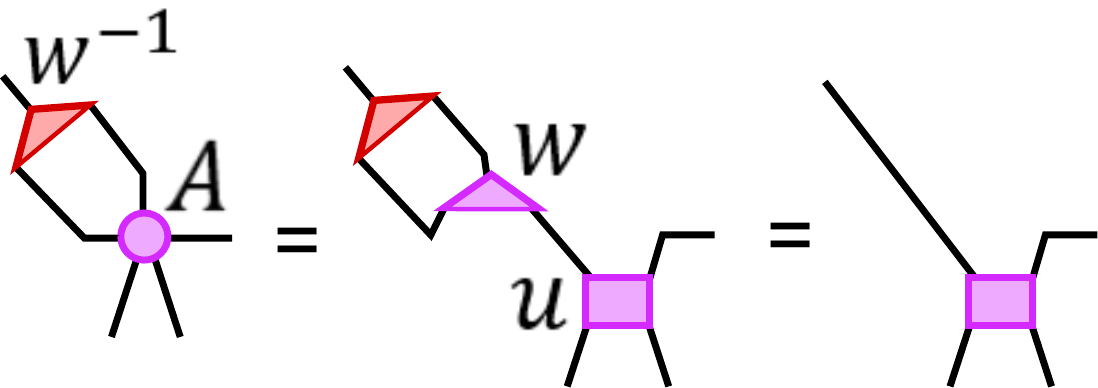}}}
\end{equation}
Since $A$ and $w^{-1}$ are symmetric, $u$ must also be symmetric fulfilling:
\begin{equation}
\vcenter{\hbox{\includegraphics[width=1.5cm]{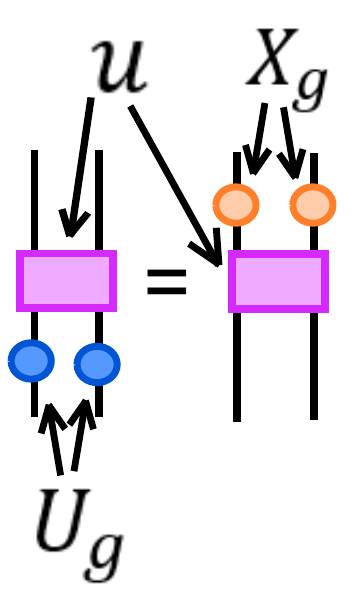}}}
\end{equation}


\begin{thebibliography}{74}
\bibitem{gauge} One can always conjugate MPS matrices $B_i$ (here, $i$ is the open index of the MPS tensor $B$) with a unitary matrix $U$, $B'_i \equiv UB_iU^\dagger$, without changing the state, while also preserving normality and the canonical form. See e.g. Theorem 7 in Ref.~\onlinecite{mps3}. However, if tensor $B$ is symmetric one can choose a $U$ such that $B'$ is not symmetric. This means that \eref{eq:mpsSym} holds only up conjugation by a (possibly non-symmetric) unitary matrix.
\bibitem{cocycle} But the phases $e^{i\omega(g,h)}$ must satisfy the so called 2-co-cycle equation, a constraint resulting from the associativity of the group product.
\bibitem{projrep} A projective representation $V_g$ of a group $\mathcal{G}$ is defined only up to a phase, $V_g \leftrightarrow e^{i\phi_g} V_g$ and $V_gV_h = e^{i\omega(g,h)}V_{gh}$, which results in equivalence classes of projective representations under the relation $\omega(g,h) \sim \omega(g,h) + \phi_g + \phi_h - \phi_{g.h}$ mod $2\pi$. The equivalence classes form a group that is isomorphic to the second cohomology group $H^2(G,U(1))$. Linear representations corresponds to identity element of $H^2(G,U(1))$. Thus, the tensor product of a projective representation and a linear representation is always a projective representation in the same equivalence class.
\bibitem{symTensorsProj} In previous works, such as Refs.~\onlinecite{symmetricTensors,symProtectedMERA}, a $\mathcal{G}$-symmetric tensor was restricted to remain invariant only under the action of linear representations on its indices. Here we will use `symmetric tensor' (without the prefix $\mathcal{G}$) to mean a tensor that remains invariant under the action of the symmetry even when the symmetry acts as a projective representation on some of the indices. Since projective representations of a group $\mathcal{G}$ are often linear representations of a larger group $\mathcal{G}'$ (e.g. a covering group of $\mathcal{G}$), a symmetric tensor that carries projective representation of $\mathcal{G}$ can also be regarded, and referred to, as a $\mathcal{G}'$-symmetric tensor.
\bibitem{optimality} These gauge transformations do not change the size of the tensors. Therfore, transforming to symmetric tensors does not result in a less optimal description of the state, that it, it does not increase the number of tensor parameters required to describe the state.
\bibitem{symcontract} We assume that the symmetric tensor $A$ is obtained by contracting tensors $u$ and $w$, which are also symmetric. Symmetric tensors $u$ and $w$ fulfill $(U_g \otimes U_g) u = u (X_g \otimes X_g)$ and $(X^\dagger_g \otimes X^\dagger_g) w = w (U'_g)^\dagger$ for all $g \in \mathcal{G}$, respectively. See also Appendix \ref{sec:appendixB}.
\end{thebibliography}
\end{document}